\documentclass[10pt,conference]{IEEEtran}
\usepackage{xurl}
\usepackage{cite}
\usepackage{amsmath,amssymb,amsfonts}
\usepackage{algorithmic}
\usepackage{graphicx}
\usepackage{textcomp}
\usepackage{xcolor}
\usepackage{enumitem}
\usepackage{hyperref}
\usepackage{titlesec}
\begin{document}

\pagestyle{plain}
\title{Evaluating GAN-LSTM for Smart Meter Anomaly Detection in Power Systems}

\author{
\IEEEauthorblockN{
Fahimeh Orvati Nia\IEEEauthorrefmark{1},
Shima Salehi,
Joshua Peeples
}
\IEEEauthorblockA{
Department of Electrical and Computer Engineering,\\
Texas A\&M University, College Station, TX, USA \\
\IEEEauthorrefmark{1}Email: fahimehorvatinia@tamu.edu
}
}
\maketitle
\begin{abstract}
Advanced metering infrastructure (AMI) provides high-resolution electricity consumption data that can enhance monitoring, diagnosis, and decision-making in modern power distribution systems. Detecting anomalies in these time-series measurements is challenging due to nonlinear, nonstationary, and multi-scale temporal behavior across diverse building types and operating conditions. This work presents a systematic, power system oriented evaluation of a GAN-LSTM framework for smart-meter anomaly detection using the Large-scale Energy Anomaly Detection (LEAD) dataset, which contains one year of hourly measurements from 406 buildings. The proposed pipeline applies consistent preprocessing, temporal windowing, and threshold selection across all methods, and compares the GAN-LSTM approach against six widely used baselines, including statistical, kernel-based, reconstruction-based, and GAN-based models. Experimental results demonstrate that the GAN-LSTM significantly improves detection performance, achieving an F1-score of 0.89. These findings highlight the potential of adversarial temporal modeling as a practical tool for supporting asset monitoring, non-technical loss detection, and situational awareness in real-world power distribution networks. The code for this work is publicly available\footnote{\href{https://github.com/fahimehorvatinia/GAN-LSTM-Smart-Meter-Anomaly-Detection}{github.com/fahimehorvatinia/GAN-LSTM-Smart-Meter-Anomaly-Detection}}.
\end{abstract}


\section{Introduction}

Modern power distribution networks are undergoing a rapid transformation due to the increasing penetration of distributed energy resources (DERs), electric vehicles, and renewable generation. While these technologies contribute to decarbonization goals, they also introduce new operational challenges, including voltage instabilities, power quality degradation, line congestion, and increased technical losses \cite{Caballero2022DERReview}. Moreover, uncoordinated or abnormal electricity consumption at the building level can propagate adverse effects throughout the distribution network, leading to inefficient operation and increased stress on grid infrastructure. At the same time, the growing deployment of sensing technologies such as phasor measurement units (PMUs), waveform recorders, and advanced metering infrastructure (AMI) has significantly improved observability in modern power systems \cite{Badawi2024POW}. Among these technologies, smart meters have become one of the most widespread sensing devices in distribution networks, providing high-resolution electricity consumption data at the building or customer level. This data enables utilities to better understand load dynamics and opens new opportunities for data-driven monitoring, diagnostics, and intelligent decision support for modern grid management.

Advanced metering infrastructure has therefore transformed the power grid into a data-rich cyber-physical system, enabling utilities to collect massive volumes of consumption time-series data that support applications such as demand forecasting, load profiling, non-technical loss detection, outage analysis, and cyber-physical security monitoring \cite{BRahaman2024MLSmartGrid}. At the same time, anomalies in smart-meter data may arise from a wide range of causes, including equipment degradation, HVAC system malfunctions, unauthorized energy usage, metering faults, or even cyber-physical intrusions \cite{Zhang2021TimeSeriesGrid}. Prompt detection of these irregular patterns is crucial for asset management, infrastructure protection, and overall grid resiliency. However, building-level electricity consumption is highly non-stationary, exhibits strong daily and seasonal variability, and differs significantly across building types and usage profiles, making anomaly detection in real-world smart-meter data a challenging task for traditional analytical methods.

Conventional anomaly detection approaches, including statistical thresholding, clustering, and classical machine learning methods, often assume simplified data distributions and struggle to capture long-term temporal dependencies \cite{Guato2024SmartGridReview, Patrizi2024PowerQuality}. Although deep learning models such as recurrent autoencoders and variational autoencoders can better model temporal dependencies, they may still fail to detect subtle or evolving anomalies that do not significantly influence point-wise reconstruction error \cite{Neloy2024AutoencodersAD}. This limitation reduces their effectiveness in practical grid monitoring scenarios.

Generative Adversarial Networks (GANs) provide an alternative perspective by modeling the underlying data distribution through an adversarial learning process \cite{esteban2017real}. When combined with Long Short-Term Memory (LSTM) networks, GANs can capture nonlinear temporal dynamics and complex consumption patterns in electricity time series. In this study, a GAN–LSTM framework is systematically evaluated for smart-meter anomaly detection under a unified preprocessing, training, and evaluation pipeline tailored to real-world power system data.

The main contributions of this paper are summarized as follows:

\begin{itemize}
    \item A power-system–oriented evaluation of a GAN–LSTM framework for modeling normal electricity consumption behavior and detecting anomalies in AMI data.
    \item A comprehensive benchmarking study against six widely used anomaly detection baselines, including statistical, kernel-based, reconstruction-based, and GAN-based methods.
    \item An experimental investigation on the Large-scale Energy Anomaly Detection (LEAD) dataset \cite{energy-anomaly-detection}, containing smart-meter data from 406 buildings and approximately 1.8 million labeled time-series windows reflective of real distribution system operations.
    \item Quantitative and qualitative analyses demonstrating the potential of adversarial temporal modeling to enhance situational awareness, support preventive maintenance, and improve reliability in modern power distribution networks.
\end{itemize}

To the best of our knowledge, this work presents the first systematic power-system-focused evaluation of a GAN–LSTM framework on the LEAD dataset. The results highlight its practical value for utilities and grid operators as a scalable tool for improving anomaly detection, enhancing operational awareness, and supporting data-driven decision making in modern distribution systems.

\section{Related Work}

Anomaly detection in electricity consumption data has been studied using statistical methods, classical machine learning, and deep learning. Early rule-based thresholds, clustering, and distance-based methods often fail when faced with the nonlinear and nonstationary behavior of smart-meter data. Classical models such as Isolation Forest \cite{Liu2008IsolationForest} and One-Class Support Vector Machine (OC-SVM) \cite{Scholkopf2001OCSVM} provide more flexible decision boundaries, but they ignore temporal dependencies and often underperform on complex time-series signals.

Deep learning methods explicitly model sequential structure. Long Short-Term Memory (LSTM) autoencoders \cite{Malhotra2015LongST} reconstruct normal patterns and flag large reconstruction errors as anomalies. Variational Autoencoders (VAEs) \cite{Kingma2013VAE} add probabilistic latent modeling to capture uncertainty in the representation. However, reconstruction-based approaches can still miss subtle or long-range deviations that do not strongly affect point-wise reconstruction error.

Generative Adversarial Networks (GANs) offer an alternative by learning data distributions through adversarial training. AnoGAN \cite{Schlegl2017UnsupervisedAD} first demonstrated GAN-based anomaly detection for images using latent-space optimization. MAD-GAN \cite{Li2019MADGAN} and TAnoGAN \cite{Bashar2020TAnoGAN} extended this idea to multivariate time series, and later models such as TimeGAN \cite{Yoon2019TimeGAN} focused on realistic sequence generation and representation learning. Most of these studies target image or generic time-series benchmarks rather than smart-meter data in power systems.

\subsection{Relation to GAN-based Time-Series Models}

The GAN-LSTM configuration evaluated in this work is most closely related to MAD-GAN and TAnoGAN. MAD-GAN uses LSTM-based generators and discriminators for multivariate time-series anomaly detection. TAnoGAN adapts AnoGAN to sequential data by combining a recurrent generator with latent-space optimization. In contrast, The analysis focuses on univariate hourly electricity consumption for individual buildings in the LEAD dataset and employs a stacked LSTM-based GAN under a fixed 60-hour windowing scheme.

Our anomaly score follows an AnoGAN-style latent inversion and combines residual and discriminator feature losses with a fixed trade-off parameter $\lambda = 0.1$. The resulting score is thresholded using a consistent procedure across all evaluated methods. TAnoGAN is implemented following \cite{Bashar2020TAnoGAN} and adapted to the same univariate LEAD setting to provide a direct GAN-based baseline. Rather than introducing a new architecture, The GAN-LSTM model is treated as a representative adversarial temporal approach and is systematically compared with statistical, kernel-based, reconstruction-based, and GAN-based anomaly detectors under a unified preprocessing and evaluation pipeline for smart-meter data.

\subsection{Dataset Structure}

This study uses the Large-scale Energy Anomaly Detection (LEAD) dataset released on Kaggle \cite{energy-anomaly-detection}. The dataset contains one year of hourly electricity consumption for 406 buildings, with 200 buildings provided for training and 206 held out for testing. Each building contributes 8{,}760 hourly measurements, which yields approximately 1.75 million labeled time steps in the training portion.

For model training, windows are constructed only from segments in which all hourly samples are labeled as normal, ensuring that the models learn the distribution of typical consumption patterns. Any window containing at least one anomalous time step is excluded from training and used only for validation or testing. From the pool of normal training windows, $10\%$ are randomly held out for validation, and all windows from the $206$ test buildings are used exclusively for final evaluation.

Each hourly record includes a building identifier, timestamp, electricity consumption in kilowatt-hours, and a binary anomaly label indicating normal or abnormal usage. The dataset also provides engineered time features, building characteristics, and weather variables. To maintain a controlled and comparable evaluation setting for temporal anomaly detection models, all experiments in this work are conducted solely on the raw consumption time series.

\subsection{Anomaly Characteristics}

Anomalies in the LEAD dataset arise from several real-world patterns:

\begin{itemize}
    \item \textbf{Sudden spikes or drops:} Caused by mechanical faults, abrupt load activation, or metering errors.
    \item \textbf{Uncharacteristic deviations:} Consumption that is inconsistent with historical behavior or seasonal expectations.
    \item \textbf{Persistent anomalies:} Long-term abnormal behavior that indicates system malfunction or misconfiguration.
    \item \textbf{Irregular fluctuations:} Rapid oscillations that violate typical load cycling characteristics, for example HVAC cycling.
\end{itemize}

Because the dataset is derived from real building telemetry, anomalies reflect environmentally driven patterns, operational changes, and occasional sensor reliability issues, which together provide a realistic and challenging benchmark for anomaly detection in smart-meter data.

\section{Method}

This section describes the complete pipeline for time-series anomaly detection in building-level smart-meter data. The workflow consists of three main stages: data preprocessing and temporal windowing, adversarial training of a GAN-LSTM model using only normal samples, and test-time anomaly detection via latent-space optimization and threshold-based decision making.

\subsection{Data Preprocessing and Temporal Windowing}

Raw hourly meter readings are transformed into fixed-length sequences suitable for sequential models. The preprocessing pipeline is designed to preserve temporal structure, stabilize building-level normalization, and ensure clear separation between training and testing data.

\textbf{Missing Value Handling.}
For each building, missing values caused by communication failures or sensor faults are imputed using forward fill along the time axis, followed by backward fill for any remaining gaps. After normalization, any isolated missing values, which are rare, are set to zero to prevent propagation of missing-value artifacts.

\textbf{Normalization.}
To prevent buildings with higher absolute consumption from dominating training, each building is standardized using a per-building z-score:
\begin{equation}
x_{\text{norm}} = \frac{x - \mu_b}{\sigma_b},
\end{equation}
where $\mu_b$ and $\sigma_b$ denote the mean and standard deviation of the annual consumption profile of building $b$. This strategy avoids data leakage across buildings and preserves relative consumption patterns.

\textbf{Temporal Windowing.}
The normalized time series is segmented into fixed-length windows using a sliding window approach. A window length of 60 hours is selected because it captures daily periodicity of 24 hours, multi-day cycles of approximately 48 to 72 hours, and characteristic HVAC operation, while remaining computationally efficient.

For training, a stride of one hour is used, yielding
\begin{equation}
L_{\text{train}} = 8{,}760 - 60 + 1 = 8{,}701
\end{equation}
overlapping windows per building. Only windows labeled as normal are used for training.

For testing, a non-overlapping stride of 60 hours is applied, resulting in
\begin{equation}
L_{\text{test}} = \left\lfloor \frac{8{,}760}{60} \right\rfloor = 146
\end{equation}
windows per building and approximately $206 \times 146 = 30{,}076$ total test sequences.

\begin{figure*}[t]
\centering
\includegraphics[width=0.95\linewidth]{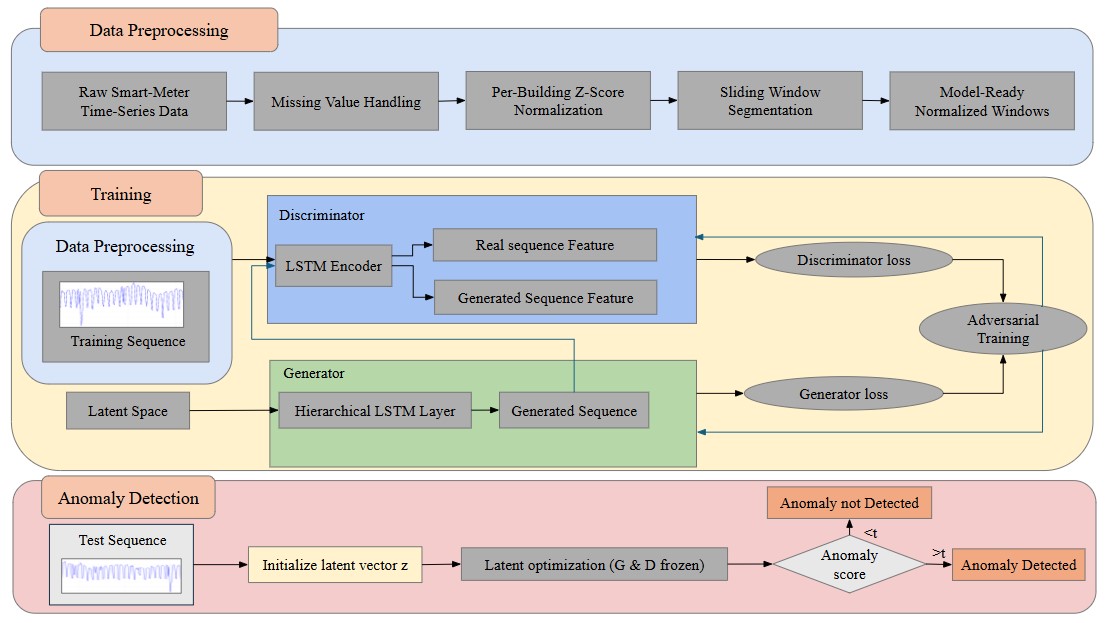}
\caption{Overview of the GAN-LSTM anomaly detection pipeline. Top: preprocessing and windowing. Center: adversarial training of the LSTM-based generator and discriminator on normal windows only. Bottom: test-time latent-space optimization with frozen networks to compute anomaly scores and final classification.}
\label{fig:gan_lstm_pipeline}
\end{figure*}

\textbf{Window-Level Labeling.}
Each 60-hour window is labeled as anomalous if at least one time step inside the window is labeled as anomalous in the LEAD ground truth. Otherwise, the window is considered normal. This rule is applied consistently to validation and test data. All evaluation metrics are computed at the window level. The overall pipeline is illustrated in Fig.~\ref{fig:gan_lstm_pipeline}.

\subsection{Model Configuration}

The GAN-LSTM follows a standard adversarial framework adapted for sequential data. The generator synthesizes realistic consumption sequences, while the discriminator distinguishes real from generated sequences. Both networks use LSTM layers to capture short- and medium-range temporal dependencies.

\textbf{Generator Architecture.}
The generator receives a latent vector $z \sim \mathcal{N}(0, 0.1^2)$ of dimension 100 and processes it through three stacked LSTM layers with 32, 64, and 128 hidden units. The final output is mapped to a 60-step sequence using a fully connected layer with Tanh activation, enabling the model to represent both local fluctuations and multi-day patterns.

\textbf{Discriminator Architecture.}
The discriminator takes a normalized 60-step input sequence and processes it with a single LSTM layer of 100 hidden units, followed by a dense sigmoid layer producing
\begin{equation}
D(x) \in [0,1],
\end{equation}
which indicates the probability that the sequence is real. Intermediate LSTM feature representations $f_D(x)$ are retained for anomaly scoring during reconstruction.

\subsection{Training Procedure}

Only windows labeled as normal from the 200 training buildings are used during adversarial training, allowing the model to learn the distribution of typical consumption behavior. Anomaly labels are not used during training and are reserved for validation and evaluation.

\textbf{Adversarial Learning Framework.}
Training alternates between discriminator and generator updates. The discriminator learns to separate real windows from generated ones. The generator is optimized to produce sequences that the discriminator classifies as real, which forces it to approximate the normal data distribution.

Both networks are trained using the Adam optimizer with
\begin{equation}
\alpha = 2 \times 10^{-4}, \quad \beta_1 = 0.5, \quad \beta_2 = 0.999,
\end{equation}
and training is performed on a Quadro RTX 6000 GPU.

\textbf{Batch Construction.}
Each mini-batch contains 32 samples: 16 real windows randomly drawn from the training set and 16 synthetic windows generated from latent samples. Synthetic data are regenerated at each iteration to improve stability and reduce discriminator overfitting.

\textbf{Loss Functions.}
The discriminator is trained using binary cross-entropy
\begin{equation}
L_D = -\frac{1}{2} \mathbb{E}_{x}[\log D(x)] - \frac{1}{2} \mathbb{E}_{z}[\log(1 - D(G(z)))],
\end{equation}
while the generator minimizes
\begin{equation}
L_G = - \mathbb{E}_{z}[\log D(G(z))],
\end{equation}
encouraging it to generate realistic sequences.

\textbf{Training Duration and Stability.}
The model is trained for 20 epochs over approximately 1.75 million windows. Stability is verified by observing a stabilizing generator loss, bounded discriminator loss oscillations, and discriminator accuracy remaining between 50\% and 90\%.
\subsection{Anomaly Detection via Latent Optimization}

After training, anomalies are detected through latent-space reconstruction rather than direct encoding, since no encoder network is used.

\textbf{Latent Inversion.}
For each test window $x$, a TAnoGAN-style latent optimization strategy is adopted~\cite{Bashar2020TAnoGAN}. The latent code is initialized from the training prior, $z_0 \sim \mathcal{N}(0, 0.1^2 I)$, and refined by minimizing
\begin{equation}
z^* = \arg\min_z \Big( (1 - \lambda)\lVert x - G(z) \rVert_1 + \lambda \lVert f_D(x) - f_D(G(z)) \rVert_1 \Big),
\end{equation}
where $\lambda = 0.1$, the first term is the residual loss, and the second term is the discriminator feature loss. The latent vector $z$ is optimized using Adam (learning rate $\eta_z = 10^{-2}$) for $\Lambda$ steps, updating only $z$ while keeping $G$ and $D$ fixed; in these experiments, $\Lambda = 300$.

\textbf{Anomaly Score and Threshold.}
The final anomaly score is
\begin{equation}
S(x) = (1 - \lambda) R(x) + \lambda F(x),
\end{equation}
evaluated at the optimised code $z^*$. A threshold $\tau$ is chosen on the validation set (windows from the 200 training buildings) by maximising the F1 score; test windows with $S(x) \ge \tau$ are flagged as anomalous and the rest as normal.

\section{Evaluation Metrics and Results}
All methods are evaluated at the window level using accuracy, precision, recall, F1-score, specificity, and ROC AUC on the held-out test set, with binary predictions obtained using a global threshold selected on the validation data.
\begin{table*}[htb]
\centering
\caption{Comparison of anomaly detection methods on the LEAD energy dataset.}
\label{tab:method_comparison}
\resizebox{0.8\textwidth}{!}{
\begin{tabular}{|l|c|c|c|c|c|}
\hline
\textbf{Method} & \textbf{Accuracy} & \textbf{Precision} & \textbf{Recall} & \textbf{F1-Score} & \textbf{ROC AUC} \\
\hline
Isolation Forest \cite{Liu2008IsolationForest}          & 57.83\% & 0.54 & 0.51 & 0.52 & 0.59 \\
One-Class SVM \cite{Scholkopf2001OCSVM}                 & 60.17\% & 0.57 & 0.56 & 0.56 & 0.61 \\
LSTM Autoencoder \cite{Malhotra2015LongST}              & 65.42\% & 0.61 & 0.59 & 0.60 & 0.70 \\
Attention LSTM Autoencoder \cite{Zhou2017AnomalyDW}     & 68.93\% & 0.65 & 0.63 & 0.64 & 0.74 \\
Variational Autoencoder \cite{Kingma2013VAE}            & 67.58\% & 0.63 & 0.61 & 0.62 & 0.72 \\
TAnoGAN \cite{Bashar2020TAnoGAN}                        & 71.86\% & 0.69 & 0.66 & 0.67 & 0.77 \\
\textbf{GAN-LSTM}                                       & \textbf{89.73\%} & \textbf{0.88} & \textbf{0.89} & \textbf{0.89} & \textbf{0.83} \\
\hline
\end{tabular}
}
\end{table*}
\subsection{Performance of the GAN-LSTM Model}

On the 206 test buildings, the proposed GAN-LSTM model achieves 89.73\% accuracy, precision 0.88, recall 0.89, F1-score 0.89, specificity 0.90, and ROC AUC 0.83, indicating strong separation between normal and anomalous consumption patterns. Each model produces a continuous anomaly score per 60-hour window; a single global threshold is selected on the validation windows from the 200 training buildings by sweeping candidate values and choosing the one that maximizes the F1-score. This threshold is then fixed and applied to the test buildings.

Figure~\ref{fig:confusion_matrix} summarizes the resulting confusion matrix for GAN-LSTM, showing high true-positive and true-negative counts with comparatively few false alarms and missed anomalies, consistent with the reported metrics.

\subsection{Comparison with Baseline Methods}

The GAN-LSTM model is compared with several widely used anomaly detection baselines: Isolation Forest~\cite{Liu2008IsolationForest}, One-Class SVM~\cite{Scholkopf2001OCSVM}, an LSTM Autoencoder~\cite{Malhotra2015LongST}, an Attention-Enhanced LSTM Autoencoder~\cite{Zhou2017AnomalyDW}, a Variational Autoencoder (VAE)~\cite{Kingma2013VAE}, and TAnoGAN~\cite{Bashar2020TAnoGAN}. All models are trained on the same 60-hour windows and evaluated on the identical test set.

\textbf{Baseline configuration.}
Isolation Forest and One-Class SVM use scikit-learn implementations with default settings, except for the contamination parameter (Isolation Forest) and $\nu$ (One-Class SVM), which are selected on the validation set to maximize F1. The LSTM Autoencoder, Attention LSTM Autoencoder, and VAE share the same 3-layer LSTM backbone as the GAN-LSTM generator and are trained with Adam under the same schedule as the proposed model. TAnoGAN follows the public implementation in~\cite{Bashar2020TAnoGAN}, adapted to the LEAD windows with the same sequence length and latent dimension as GAN-LSTM.

As shown in Table~\ref{tab:method_comparison}, GAN-LSTM substantially outperforms all baselines across all metrics. Classical tree- and kernel-based methods struggle to capture complex temporal structure, while reconstruction-based deep models and TAnoGAN perform better but still fall short of the proposed adversarial temporal model. Although GAN-LSTM achieves high accuracy and F1-score at the selected operating point, the more conservative ROC AUC reflects the challenge of separating borderline anomalous and normal windows across the full decision spectrum.

\subsection{Visualization of Model Performance}

\begin{figure}[htbp]
\centering
\includegraphics[width=0.45\textwidth]{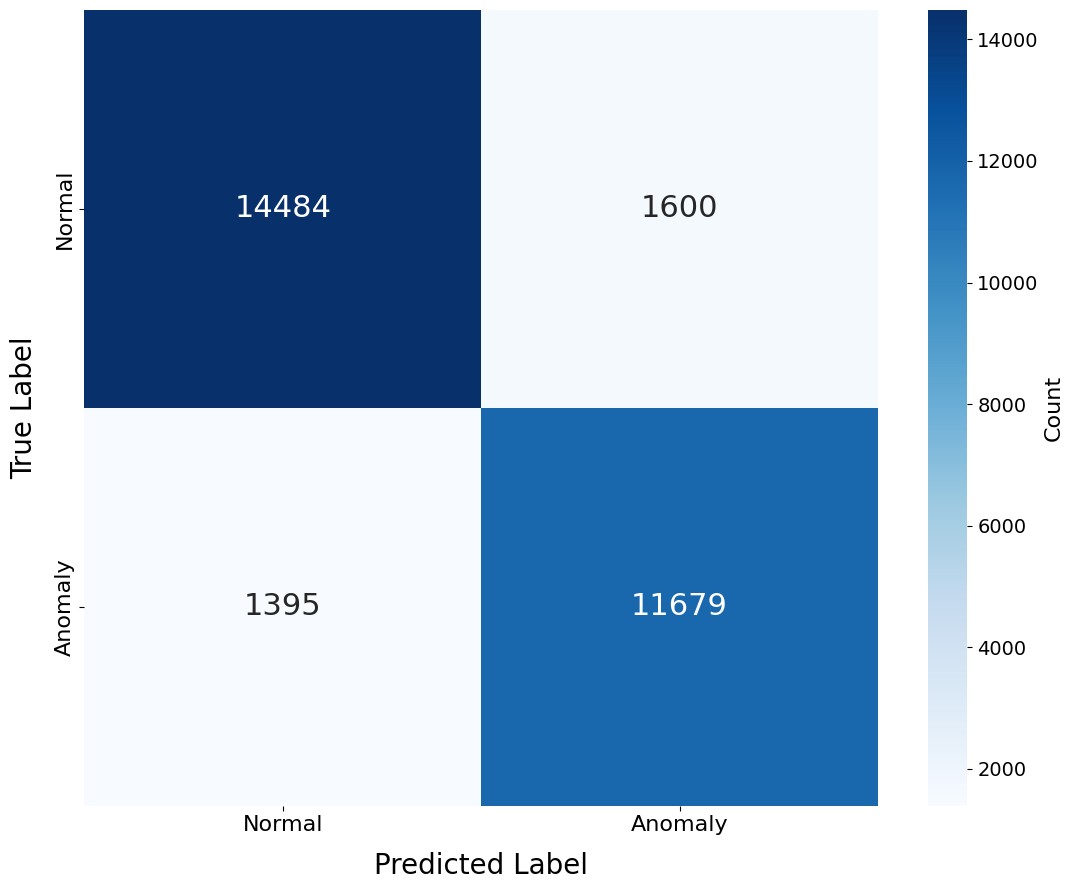}
\caption{Confusion matrix for the GAN-LSTM model on the test windows.}
\label{fig:confusion_matrix}
\end{figure}

\begin{figure}[htbp]
\centering
\includegraphics[width=0.48\textwidth]{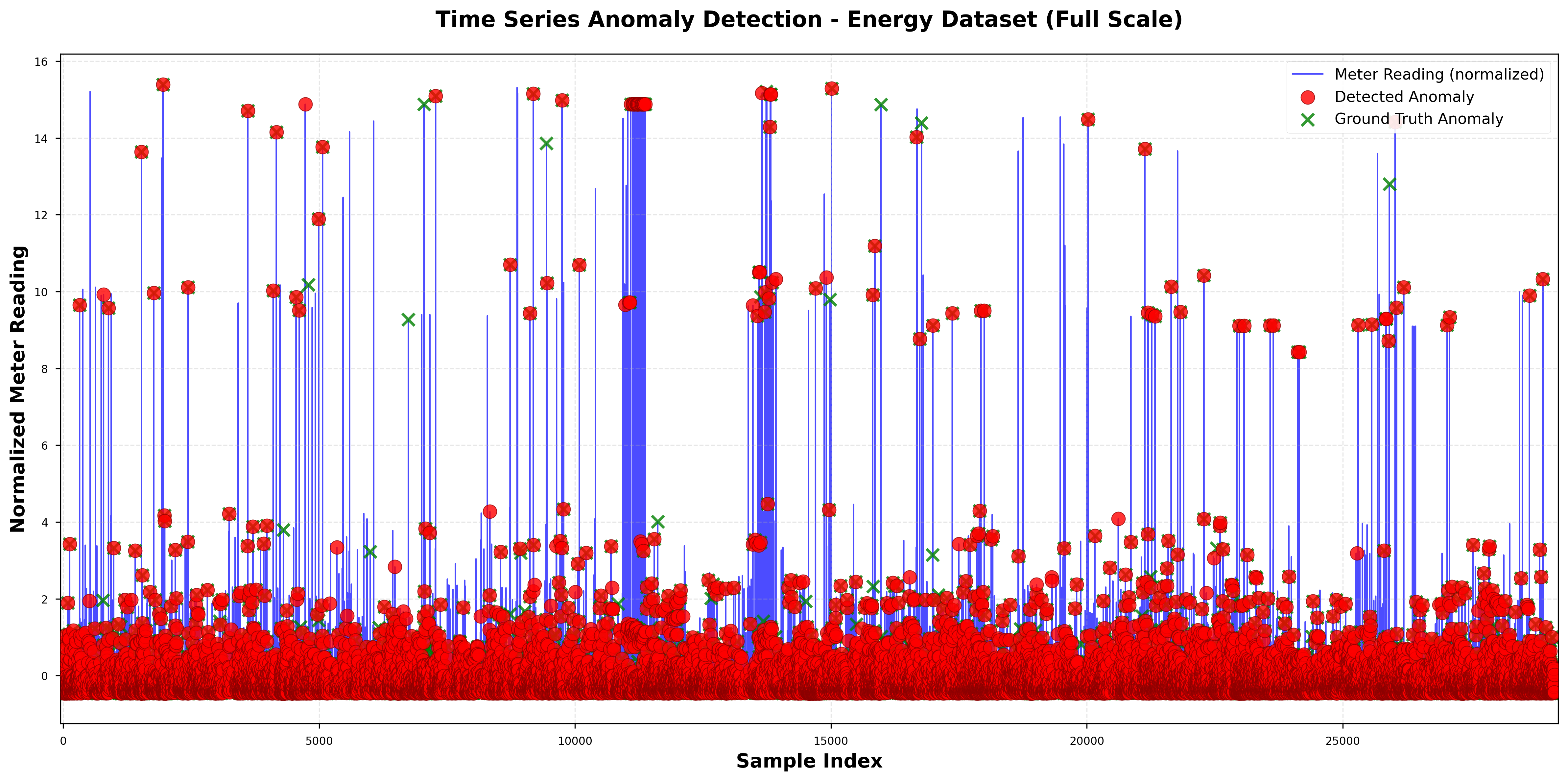}
\caption{Full-year consumption sequence for a sample building with detected anomalies (red) and ground-truth anomalies (green).}
\label{fig:ts_full}
\end{figure}

\begin{figure}[htbp]
\centering
\includegraphics[width=0.48\textwidth]{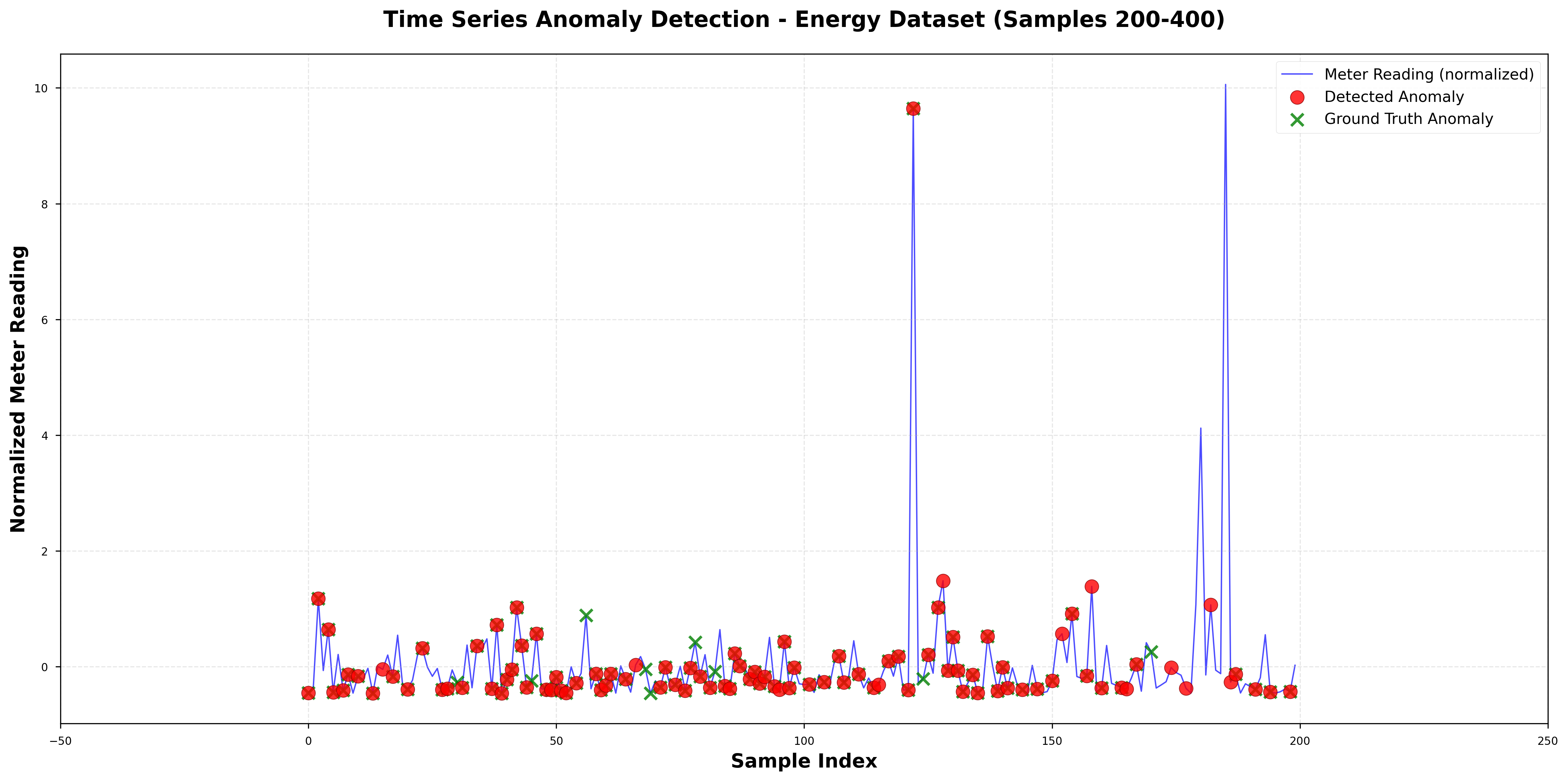}
\caption{Zoomed view of samples 200--400 showing anomaly detections in a dense irregular region.}
\label{fig:ts_zoom2}
\end{figure}

Figure~\ref{fig:confusion_matrix} shows the confusion matrix for GAN-LSTM on the test windows. To illustrate temporal behavior, Fig.~\ref{fig:ts_full} plots a full-year consumption profile for a sample building with detected anomalies overlaid on the ground-truth labels, and Fig.~\ref{fig:ts_zoom2} zooms into a dense anomalous region, showing that the model tracks both sharp irregularities and more gradual deviations.

\subsection{Discussion}

Overall, GAN-LSTM achieves higher accuracy, F1-score, and ROC AUC than classical statistical models, deep reconstruction architectures, and prior GAN-based approaches. The hierarchical LSTM generator and discriminator, combined with adversarial training and latent-space optimisation, provide a richer model of normal consumption sequences and improve detection of deviations compared with shallower recurrent GANs such as TAnoGAN.

This study focuses on univariate electricity consumption to isolate the impact of temporal sequence modelling, despite the availability of metadata such as weather, calendar, and building characteristics. Extending GAN-LSTM and the baselines to multivariate inputs with exogenous covariates is a natural direction for future work. Limitations include occasional false alarms on rare but legitimate behaviours, reduced robustness under large regime shifts (e.g., extreme weather), and the computational cost of latent-space optimisation at inference time; addressing these through improved handling of nonstationarity, incorporation of contextual features, and more lightweight inference is left for future work.

\section{Conclusion}

This paper presented a systematic evaluation of a GAN--LSTM framework for anomaly detection in building-level electricity consumption data. By integrating LSTM layers into both the generator and discriminator, the proposed model effectively captures nonlinear and long-range temporal dependencies that are characteristic of real-world load behavior in modern distribution systems. Experimental results on the LEAD dataset demonstrate substantial performance improvements over classical statistical models and deep reconstruction-based methods, including Isolation Forest, One-Class SVM, LSTM and attention-based autoencoders, VAE, and TAnoGAN. The proposed approach achieved 89.73\% accuracy, 0.88 precision, 0.89 recall, and a ROC AUC of 0.83, highlighting its ability to accurately identify both isolated and persistent anomalies in smart-meter data.

From a power-system perspective, these results underscore the potential of adversarial temporal modeling as a practical tool for enhancing situational awareness in advanced metering infrastructure. The proposed framework can support utilities in identifying abnormal consumption patterns associated with equipment faults, inefficient operation, unauthorized usage, or cyber--physical threats. By enabling earlier and more accurate detection of such events, the method offers practical value for improving grid reliability, reducing technical losses, and supporting data-driven decision-making in modern distribution networks. Future work will focus on extending the framework to multivariate and spatio-temporal\cite{nia2025traffic} settings by incorporating additional contextual information such as weather data, calendar effects, and building characteristics. Recent advances in spatio-temporal modeling have shown strong potential for capturing complex dynamic behaviors, motivating the exploration of such architectures for large-scale energy monitoring and forecasting applications. Additional efforts will target improvements in inference efficiency and robustness to nonstationarity to support deployment in near-real-time monitoring systems for large distribution networks.

\bibliographystyle{ieeetr}
\bibliography{egbib}

\end{document}